\documentclass[prd,showpacs,showkeys,nofootinbib,floatfix,
               fleqn,preprint,12pt,tightenlines]{revtex4-1}

\newcommand{\version}{v3} 

\usepackage{amsmath,amssymb,revsymb,graphicx,dcolumn}
\usepackage{array,hyperref,rotating}
\usepackage{dcolumn}  

\newcommand{\beq}{\begin{equation}}
\newcommand{\eeq}{\end{equation}}
\newcommand{\beqa}{\begin{eqnarray}}
\newcommand{\eeqa}{\end{eqnarray}}
\newcommand{\bsubeqs}{\begin{subequations}}
\newcommand{\esubeqs}{\end{subequations}}

\newcommand{\half}{{\textstyle \frac{1}{2}}}

\begin{document}

\begin{widetext}
%
\noindent arXiv:2307.12876 \hfill KA--TP--15--2023\;(\version)
\newline\vspace*{0mm}
\end{widetext}

\title{Higher-dimensional extension of a vacuum-defect wormhole}

 \author{\vspace*{4mm}F.R. Klinkhamer}
 \email{frans.klinkhamer@kit.edu}
 \affiliation{Institute for Theoretical Physics,
 Karlsruhe Institute of Technology (KIT),\\
 76128 Karlsruhe, Germany\\}

\vspace*{10mm}

\begin{abstract}
\vspace*{2.5mm}\noindent
We present a 5D metric which interpolates between the
standard 4D Schwarzschild metric with mass parameter $M$
and a new 4D $M$-deformed vacuum-defect-wormhole metric.
The 5D spacetime can, in principle, have an infinite mass density
that gives rise to the $M$ parameter of the
4D $M$-deformed vacuum-defect wormhole.
For completeness, we also give 4D interpolating metrics  
between the original vacuum-defect wormhole
at the wormhole throat and the
$M$-deformed vacuum-defect wormhole at spatial infinity.
\vspace*{10mm}
\end{abstract}


\maketitle

\section{Introduction}
\label{sec:Introduction}

The traversability of 4D wormhole solutions appears to rely on
the presence of exotic
matter~\cite{Ellis1973,Bronnikov1973,MorrisThorne1988,Visser1996}.
But recently an alternative has been proposed~\cite{Klinkhamer2023-defect-WH},
with a follow-up paper in Ref.~\cite{Klinkhamer2023-mirror-world}
and a brief review in Ref.~\cite{Klinkhamer2023-review}.
The idea is that the exotic input now resides in the spacetime
structure itself and not in the matter content.
Specifically, the considered metric is degenerate, with a vanishing
metric determinant on a submanifold. This particular metric
structure has been called a ``spacetime defect,'' where the analogy
is with defects in an atomic crystal.

A special case of this new type of 4D traversable wormhole has no matter
at all and is called the vacuum-defect-wormhole solution. Wang~\cite{Wang2023}
has noticed that this vacuum-defect-wormhole metric can be deformed by the
introduction of a mass parameter $M$ entering Schwarzschild-type factors
in the metric (related work has also been reported by Ahmed~\cite{Ahmed2023}).
This 4D $M$-deformed-vacuum-defect-wormhole metric 
still solves the vacuum Einstein equation
(just as for the original Schwarzschild-spacetime-defect metric  
of Sec.~3 in Ref.~\cite{Klinkhamer2014-MPLA}).

However, the question remains as to the physical interpretation of $M$,
because the 4D manifold is geodesically complete and there appears to be no
place for a point mass (the original 4D Schwarzschild
spacetime~\cite{Schwarzschild1916,Painleve1921,Gullstrand1922}
is geodesically incomplete and there can be a point mass located
outside the manifold proper, at the center of
a 3-space with spherical symmetry).

In the Note Added of Ref.~\cite{Klinkhamer2023-review},
we have suggested that a physical
explanation of the mass parameter $M$
in the deformed-vacuum-defect-wormhole metric
may require higher dimensions.
It turns out to be not quite trivial to get an appropriate
higher-dimension metric, but we have succeeded in the end, even
though the obtained 5D metric appears somewhat baroque at first sight.

\section{Five-dimensional spacetime}
\label{sec:Five-dimensional-spacetime}

\subsection{Construction}
\label{subsec:Construction}

There are three steps for getting the desired
higher-dimensional spacetime:\vspace*{-1mm}
\begin{enumerate}
\item
start from the 4D vacuum-defect-wormhole metric
(length scale $b>0$), which is
Ricci flat and therefore, a solution of the classical
vacuum Einstein equation;\vspace*{-1mm}
\item
add a fifth dimension with radial
coordinate $\chi \in (0,\,\infty)$;\vspace*{-1mm}
\item
introduce the deformation parameter $M >0$.
\end{enumerate}

Specifically we consider a 5-dimensional spacetime
with coordinates
\bsubeqs\label{eq:coordinates}
\beqa
t &\in& (-\infty,\,\infty)\,,
\eeqa       
\beqa       
\chi &\in& (0,\,\infty)\,,
 \\[1mm]
\xi &\in& (-\infty,\,\infty)\,,\quad
\theta \in [0,\,\pi]\,,\quad
\phi \in [0,\,2\pi)\,,
\eeqa
\esubeqs
and a 5D metric from the following \emph{Ansatz}
(in units with $c=1$ and $G=1$):
\bsubeqs\label{eq:5D-metric}
\begin{eqnarray}
\label{eq:5D-metric-ds2}
ds^{2}
&\equiv&
g_{\mu\nu}(x)\, dx^\mu\,dx^\nu
\nonumber\\[1mm]
&=&
- dt^{2} +
\left(d\chi + \sqrt{\frac{2M}{\chi\,\sqrt{1+\xi^{2}/b^{2}}}}\,dt\right)^{2}
+
\left(1-\frac{2M}{\sqrt{b^{2}+\xi^{2}}}\right)^{-1}\,
\frac{\xi^{2}}{b^{2}+\xi^{2}}\;d\xi^{2}
\nonumber\\[1mm]
&&
+ \left(b^{2} +\frac{M}{M+\delta M}\,\left(\chi^2-b^2\right) + \xi^{2}\right)\,
  \Big[ d\theta^{2} + \sin^{2}\theta\, d\phi^{2} \Big]\,,
\\[2mm]
b &>& 0\,,
\\[2mm]
M &\in& \big[0,\,b/2\big)\,,
\\[2mm]
\delta M &=& 0^{+}\,,
\end{eqnarray}
\esubeqs
where the ratio $M/(M+\delta M)$ in 
\eqref{eq:5D-metric-ds2}
takes the value $0$ for $M =0$  and $1$ for $M > 0$.
The \textit{raison d'$\:\hat{\!\text{e}}$tre} of metric 
\eqref{eq:5D-metric} will become clear if
we consider certain limiting cases.

\subsection{4D submanifolds}
\label{subsec:4D-submanifolds}

For fixed $\chi=b$ and $M=0$, we recover
from \eqref{eq:5D-metric}
the 4D vacuum-defect-wormhole metric~\cite{Klinkhamer2023-defect-WH}:
\begin{eqnarray}
\label{eq:4D-metric-vacuum-defect-wormhole}
&&ds^{2}\,\Big|^{\chi=b}_{M =0}
=
- dt^{2}
+ \frac{\xi^{2}}{b^{2}+\xi^{2}}\;d\xi^{2}
+ \left(b^{2} + \xi^{2}\right)\,
  \Big[ d\theta^{2} + \sin^{2}\theta\, d\phi^{2} \Big]\,.
\end{eqnarray}
The full 5D metric for this mass parameter ($M=0$)
is, in fact, $\chi$ independent, so that the
Ricci curvature scalar $R\equiv g^{\mu\nu}\,R_{\mu\nu}$
and the Kretschmann curvature
scalar $K\equiv R^{\mu\nu\rho\sigma}\,R_{\mu\nu\rho\sigma}$
vanish over the whole 5D manifold.

For fixed $\chi=b$ and $M >0$,
we recover  the 4D $M$-deformed-vacuum-defect-wormhole metric
of Wang~\cite{Wang2023} (see also Ahmed~\cite{Ahmed2023}):
\begin{eqnarray}
\label{eq:4D-metricSchwarzschild-defect-wormhole}
&&ds^{2}\,\Big|^{\chi=b}_{M >0}
=
- \left(1-\frac{2M}{\sqrt{b^{2}+\xi^{2}}}\right)\;dt^{2}
\nonumber\\[1mm]
&&
+
\left(1-\frac{2M}{\sqrt{b^{2}+\xi^{2}}}\right)^{-1}\,
\frac{\xi^{2}}{b^{2}+\xi^{2}}\;d\xi^{2}
+ \left(b^{2} + \xi^{2}\right)\,
  \Big[ d\theta^{2} + \sin^{2}\theta\, d\phi^{2} \Big]\,,
\end{eqnarray}
which requires $M<b/2$ due to the Schwarzschild-type coordinates
used (this point will be discussed later, in an appendix).  
The full 5D metric for this mass parameter ($M>0$)
is now $\chi$ dependent, so that the Ricci curvature scalar $R$
is nonvanishing over the whole 5D manifold, also on the
$\chi=b$ slice.

\begin{figure}[t]   
\vspace*{-0mm}
\begin{center}
\includegraphics[width=0.60\textwidth]
{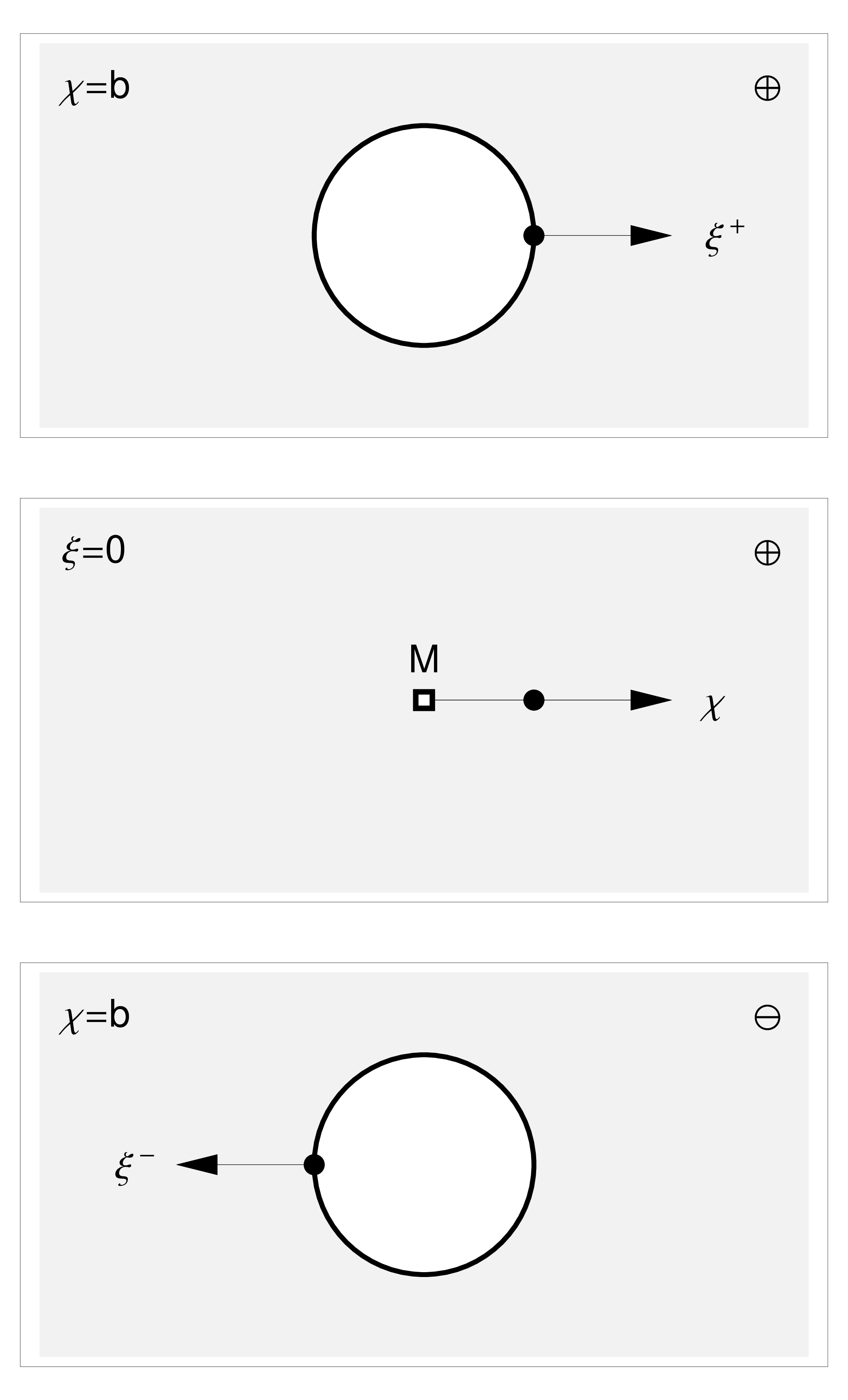}
\end{center}
\vspace*{-0mm}
\caption{Three slices of the 5D spacetime
from \eqref{eq:5D-metric}.
Each slice has $\theta=\pi/2$ and equal $t=\text{const}$.
The middle panel is for $\xi=0$ and the top and
bottom panels for $\chi=b$.
The heavy dots are identified and correspond to a single
spacetime point.
In each panel, there is an azimuthal angle $\phi \in [0,\,2\pi)$,
giving $2\pi b$ for the circumferences of the
holes in the top and bottom panels,
which have handedness $+/-$
as indicated in the top-right corner
(handedness $+$ has been chosen for the middle panel).   
There are also outgoing radial coordinates:
$\chi \geq 0$ in the middle panel,
$\xi^{+} \equiv \xi \geq 0$  for $\xi \geq 0$ in the top panel,
and $\xi^{-} \equiv -\xi \geq 0$  for $\xi \leq 0$ in the bottom panel.
}
\label{fig:three-slices of-5D-spacetime}
\vspace*{00mm}
\end{figure}

For fixed $\xi=0$ and $M >0$,
we recover the 4D Schwarzschild metric~\cite{Schwarzschild1916} in
Painlev\'{e}--Gullstrand coordinates~\cite{Painleve1921,Gullstrand1922}:
\begin{eqnarray}
\label{eq:4D-metric-Schwarzschild-PG-coordinates}
&&ds^{2}\,\Big|^{\xi=0}_{M>0}
=
- dt^{2} +
\left(d\chi + \sqrt{\frac{2M}{\chi\,}}\,dt\right)^{2}
+ \chi^2\,  \Big[ d\theta^{2} + \sin^{2}\theta\, d\phi^{2} \Big]\,.
\end{eqnarray}
The Painlev\'{e}--Gullstrand coordinates remove the
apparent Schwarzschild singularity at $\chi_{S} \equiv 2M$
and are crucial for our construction
(introducing $\xi$ will not introduce a spurious
singularity at $\chi_{S}$).
In a way, the $\xi=0$ Schwarzschild
submanifold from \eqref{eq:4D-metric-Schwarzschild-PG-coordinates}
sits between the two sheets of the
$M$-deformed-vacuum-defect wormhole
from \eqref{eq:4D-metricSchwarzschild-defect-wormhole};
see Fig.~\ref{fig:three-slices of-5D-spacetime} for a sketch.
The full 5D metric for $M>0$ is both $\chi$ and $\xi$ dependent,
so that the Ricci curvature scalar $R$
is nonvanishing over the whole 5D manifold, also on the $\xi=0$ slice.

\section{5D curvature and physical interpretation}
\label{sec:5D-curvature-and-physical-interpretation}

The curvature scalars $R$ and $K$
from the 5D metric \eqref{eq:5D-metric}, for $b=1$ and $M=1/4$,
can be calculated analytically and are given by
the following expressions:
\bsubeqs\label{eq:R-and-K-b1-Mquarter}
\beqa  
\label{eq:R-b1-Mquarter}
R\,\Big|^{b=1}_{M=1/4}
&=&
\frac{1}{32\,{\chi}^3\,
    {\left( 1 + {\xi}^2 \right) }^{5/2}\,
    {\left( {\chi}^2 + {\xi}^2 \right) }^2}\;
\Big[32\,{\xi}^4\,{\left( 1 + {\xi}^2 \right) }^2 +
    {\chi}^6\,\left( 2 + 2\,{\xi}^2 - {\sqrt{1 + {\xi}^2}} \right)
\nonumber\\[1.000mm]
&&
    - 32\,{\chi}^5\,{\left( 1 + {\xi}^2 \right) }^2\,
     \left( -1 + 4\,{\sqrt{1 + {\xi}^2}} \right)
    + 2\,{\chi}^4\,{\xi}^2\,\left( 2 + 2\,{\xi}^2 - {\sqrt{1 + {\xi}^2}} \right)
\nonumber\\[1.000mm]
&&
     -  32\,{\chi}^3\,{\left( 1 + {\xi}^2 \right) }^2\,
     \left( 1 - 2\,{\sqrt{1 + {\xi}^2}} + 4\,{\xi}^2\,{\sqrt{1 + {\xi}^2}}
       \right)
\nonumber\\[1.000mm]
&&
       +
        {\chi}^2\,{\xi}^2\,
     \left( 64 + 66\,{\xi}^4 - {\xi}^2\,
        \left( -130 + {\sqrt{1 + {\xi}^2}} \right)  \right) \Big]\,,
\\[0.75mm] 
\label{eq:K-b1-Mquarter}
K\,\Big|^{b=1}_{M=1/4}
&=&
\frac{1}
    {1024\,{\chi}^6\,{\left( 1 + {\xi}^2 \right) }^{9/2}\,
    {\left( {\chi}^2 + {\xi}^2 \right) }^4}\;
\Big[
1024\,{\xi}^8\,{\left( 1 + {\xi}^2 \right) }^{7/2}+ \dots
\nonumber\\[1.000mm]
&&
+
64\,{\chi}^{13}\,\left( 1 + {\xi}^2 \right) \,
  \left( -13 - 9\,{\xi}^2 + 12\,{\sqrt{1 + {\xi}^2}} \right)
\Big]\,,
\eeqa
\esubeqs
where the numerators have
been expanded in powers of $\chi$, with only two terms shown
for the $K$ numerator. Incidentally, the $K$ numerator at fixed
$\chi$ goes as $|\xi|^{15}$ for $\xi \to \pm\infty$
(further results for general $M>0$ are given
in App.~\ref{app:5D-curvature-scalars-for-general-pos-M}).
As mentioned before, the 5D metric \eqref{eq:5D-metric}
for $M>0$ has a nonvanishing Ricci tensor.
This implies that the vacuum Einstein equation is not solved and
an appropriate matter content is required.

From the curvature scalars in \eqref{eq:R-and-K-b1-Mquarter}
and \eqref{eq:R-and-K-b1-Mpos},
we have a manifest singularity as $\chi \to 0^{+}$ and, most likely,
the 5D manifold is geodesically incomplete.
This then answers the  question in
Sec.~\ref{sec:Introduction}, namely, what the
physical interpretation of the parameter $M$ in the
4D deformed-vacuum-defect-wormhole metric is.
In short, that parameter $M$ may trace back to a static
infinite mass density at $\chi=0$ in our 5D spacetime
(cf. Fig.~\ref{fig:three-slices of-5D-spacetime}).
In principle, this static infinite mass density
could cover (part of) a 4D submanifold
with $\chi=0$, $t \in \mathbb{R}$, $\xi \in \mathbb{R}$,
$\theta \in [0,\,\pi]$, and $\phi \in [0,\,2\pi)$.

The present paper is about higher-dimensional interpolating metrics 
(incidentally, our 5D interpolating metric is similar in spirit 
to the higher-dimensional metrics 
discussed in Ref.~\cite{DeshpandeLunin2022}).
But, for completeness, we also give, 
in App.~\ref{app:4D-interpolating-metrics}, 4D interpolating metrics 
between the original ($M=0$) vacuum-defect wormhole
at the wormhole throat and the
$M$-deformed vacuum-defect wormhole at spatial infinity.

\begin{acknowledgments}
It is a pleasure to thank Z.L. Wang for helpful discussions
on the preprint.
\end{acknowledgments}

\begin{appendix}
\section{5D curvature scalars for general positive $M$}
\label{app:5D-curvature-scalars-for-general-pos-M}

The curvature scalars
$R\equiv g^{\mu\nu}\,R_{\mu\nu}$
and $K\equiv R^{\mu\nu\rho\sigma}\,R_{\mu\nu\rho\sigma}$
from the 5D metric \eqref{eq:5D-metric}, for $b=1$ and $M>0$, are given by
\bsubeqs\label{eq:R-and-K-b1-Mpos}
\beqa
&&    
R\,\Big|^{b=1}_{M>0}
=
\frac{1}{4\,{\chi}^3\,{\left( 1 + {\xi}^2 \right) }^{5/2}\,
  {\left( {\chi}^2 + {\xi}^2 \right) }^2}\;
\nonumber\\[1.0mm]&&
\Big[
-16\,{\chi}^5\,{\left( 1 + {\xi}^2 \right) }^{5/2} -
  8\,{\chi}^3\,{\left( 1 + {\xi}^2 \right) }^{5/2}\,
   \left( -1 + 2\,{\xi}^2 \right)
\nonumber\\[1.0mm]&&
     + M\,\Big\{ {\chi}^6\,\left( 1 + {\xi}^2 \right)  -
     16\,{\chi}^3\,{\left( 1 + {\xi}^2 \right) }^2 +
     16\,{\chi}^5\,{\left( 1 + {\xi}^2 \right) }^2
\nonumber\\[1.0mm]&&
      +  16\,{\xi}^4\,{\left( 1 + {\xi}^2 \right) }^2 +
     {\chi}^4\,\left( 2\,{\xi}^2 + 2\,{\xi}^4 \right)  +
     {\chi}^2\,\left( 32\,{\xi}^2 + 65\,{\xi}^4 + 33\,{\xi}^6 \right)
      \Big\}
\nonumber\\[1.0mm]&&
   + M^2\,\Big\{ -2\,{\chi}^6\,{\sqrt{1 + {\xi}^2}} -
     4\,{\chi}^4\,{\xi}^2\,{\sqrt{1 + {\xi}^2}} -
     2\,{\chi}^2\,{\xi}^4\,{\sqrt{1 + {\xi}^2}} \Big\}
 \Big]\,,
\eeqa
\beqa
&&       
K\,\Big|^{b=1}_{M>0}
=
\frac{1}
    {16\,{\chi}^6\,{\left( 1 + {\xi}^2 \right) }^{9/2}\,
  {\left( {\chi}^2 + {\xi}^2 \right) }^4}\;
\nonumber\\[1.00mm]&&
\Big[
128\,{\chi}^{10}\,{\left( 1 + {\xi}^2 \right) }^{9/2} +
  256\,{\chi}^8\,{\xi}^2\,{\left( 1 + {\xi}^2 \right) }^{9/2}
  +   64\,{\chi}^6\,{\left( 1 + {\xi}^2 \right) }^{9/2}\,
   \left( 3 + 2\,{\xi}^4 \right)
\nonumber\\[1.00mm]&&
\eeqa
\beqa
&&
     + M\,\Big\{ -256\,{\chi}^{10}\,{\left( 1 + {\xi}^2 \right) }^4 -
     256\,{\chi}^8\,\left( -1 + {\xi}^2 \right) \,
      {\left( 1 + {\xi}^2 \right) }^4
\nonumber\\[1.00mm]&&
      -   256\,{\chi}^6\,{\left( 1 + {\xi}^2 \right) }^4\,
      \left( 3 + 2\,{\xi}^2 \right)  +
     {\chi}^{13}\,\left( -36\,\left( 1 + {\xi}^2 \right)  -
        36\,{\xi}^2\,\left( 1 + {\xi}^2 \right)  \right)
\nonumber\\[1.00mm]&&
        +  {\chi}^{11}\,\left( -36\,\left( 1 + {\xi}^2 \right)  -
        216\,{\xi}^2\,\left( 1 + {\xi}^2 \right)  -
        180\,{\xi}^4\,\left( 1 + {\xi}^2 \right)  \right)
\nonumber\\[1.00mm]&&
        + {\chi}^9\,\left( -544\,\left( 1 + {\xi}^2 \right)  -
        1712\,{\xi}^2\,\left( 1 + {\xi}^2 \right)  -
        2008\,{\xi}^4\,\left( 1 + {\xi}^2 \right)  -
        840\,{\xi}^6\,\left( 1 + {\xi}^2 \right)  \right)
\nonumber\\[1.00mm]&&
        + {\chi}^7\,\Big( -768\,\left( 1 + {\xi}^2 \right)  -
        3392\,{\xi}^2\,\left( 1 + {\xi}^2 \right)  -
        5624\,{\xi}^4\,\left( 1 + {\xi}^2 \right)  -
        4288\,{\xi}^6\,\left( 1 + {\xi}^2 \right)
\nonumber\\[1.00mm]&&
        - 1288\,{\xi}^8\,\left( 1 + {\xi}^2 \right)  \Big)
 +  {\chi}^3\,\left( -4\,{\xi}^8\,{\left( 1 + {\xi}^2 \right) }^2 -
        4\,{\xi}^{10}\,{\left( 1 + {\xi}^2 \right) }^2 \right)
\nonumber\\[1.00mm]&&
        + {\chi}^5\,\left( -544\,{\xi}^4\,\left( 1 + {\xi}^2 \right)  -
        1648\,{\xi}^6\,\left( 1 + {\xi}^2 \right)  -
        1700\,{\xi}^8\,\left( 1 + {\xi}^2 \right)  -
        596\,{\xi}^{10}\,\left( 1 + {\xi}^2 \right)  \right)\Big\}
\nonumber\\[0.50mm]
&&\nonumber
      + M^2\,\Big\{ 11\,{\chi}^{12}\,{\left( 1 + {\xi}^2 \right) }^{3/2} +
     192\,{\chi}^{13}\,{\left( 1 + {\xi}^2 \right) }^{3/2} +
     8\,{\chi}^3\,{\xi}^8\,{\left( 1 + {\xi}^2 \right) }^{5/2}
\nonumber\\[1.00mm]&&
     +  256\,{\xi}^8\,{\left( 1 + {\xi}^2 \right) }^{7/2} +
     8\,{\chi}^2\,{\xi}^6\,{\left( 1 + {\xi}^2 \right) }^{5/2}\,
      \left( 128 + 133\,{\xi}^2 \right)
\nonumber\\[1.00mm]&&
      +  32\,{\chi}^5\,{\xi}^4\,{\left( 1 + {\xi}^2 \right) }^{3/2}\,
      \left( 20 + 41\,{\xi}^2 + 27\,{\xi}^4 \right)
   +  32\,{\chi}^9\,{\left( 1 + {\xi}^2 \right) }^{3/2}\,
      \left( 36 + 77\,{\xi}^2 + 77\,{\xi}^4 \right)
\nonumber\\[1.00mm]&&
      + 4\,{\chi}^{10}\,{\sqrt{1 + {\xi}^2}}\,
      \left( 58 + 159\,{\xi}^2 + 133\,{\xi}^4 + 32\,{\xi}^6 \right)
       +  {\chi}^{11}\,\left( 72\,{\left( 1 + {\xi}^2 \right) }^{3/2} +
        840\,{\xi}^2\,{\left( 1 + {\xi}^2 \right) }^{3/2} \right)
\nonumber\\[1.00mm]&&
      + {\chi}^4\,{\xi}^4\,{\sqrt{1 + {\xi}^2}}\,
      \left( 1792 + 5536\,{\xi}^2 + 5707\,{\xi}^4 + 1963\,{\xi}^6 \right)
\nonumber\\[1.00mm]&&
      +  2\,{\chi}^8\,{\sqrt{1 + {\xi}^2}}\,
      \left( 128 + 400\,{\xi}^2 + 321\,{\xi}^4 - 79\,{\xi}^6 - 128\,{\xi}^8
        \right)
\nonumber\\[1.00mm]&&
        + 4\,{\chi}^6\,{\sqrt{1 + {\xi}^2}}\,
      \left( 192 + 1088\,{\xi}^2 + 2284\,{\xi}^4 + 2179\,{\xi}^6 +
        887\,{\xi}^8 + 96\,{\xi}^{10} \right)
\nonumber\\[1.00mm]&&
        +      {\chi}^7\,\left( 1536\,{\left( 1 + {\xi}^2 \right) }^{3/2} +
        6400\,{\xi}^2\,{\left( 1 + {\xi}^2 \right) }^{3/2} +
        8304\,{\xi}^4\,{\left( 1 + {\xi}^2 \right) }^{3/2} +
        4208\,{\xi}^6\,{\left( 1 + {\xi}^2 \right) }^{3/2} \right) \Big\}
\nonumber\\[0.50mm]
&&
   + M^3\,\Big\{ -44\,{\chi}^{12}\,\left( 1 + {\xi}^2 \right)  -
     256\,{\chi}^{13}\,\left( 1 + {\xi}^2 \right)  -
     1024\,{\chi}^{11}\,{\xi}^2\,\left( 1 + {\xi}^2 \right)
\nonumber\\[1.00mm]&&
     -  80\,{\chi}^2\,{\xi}^8\,{\left( 1 + {\xi}^2 \right) }^2 -
     128\,{\chi}^5\,{\xi}^4\,\left( 1 + {\xi}^2 \right) \,
      \left( 1 + 2\,{\xi}^2 + 3\,{\xi}^4 \right)
\nonumber\\[1.00mm]&&
      -  128\,{\chi}^9\,\left( 1 + {\xi}^2 \right) \,
      \left( 1 + 2\,{\xi}^2 + 13\,{\xi}^4 \right)  -
     16\,{\chi}^{10}\,\left( 13 + 37\,{\xi}^2 + 24\,{\xi}^4 \right)
\nonumber\\[1.00mm]&&
     - 24\,{\chi}^8\,{\xi}^2\,\left( 24 + 59\,{\xi}^2 + 35\,{\xi}^4 \right)  -
     16\,{\chi}^6\,{\xi}^4\,\left( 38 + 87\,{\xi}^2 + 49\,{\xi}^4 \right)
\nonumber\\[1.00mm]&&
     -  4\,{\chi}^4\,{\xi}^6\,\left( 80 + 171\,{\xi}^2 + 91\,{\xi}^4
     \right)
\nonumber\\[1.00mm]&&
     + {\chi}^7\,\left( -256\,{\xi}^2\,\left( 1 + {\xi}^2 \right)  -
        512\,{\xi}^4\,\left( 1 + {\xi}^2 \right)  -
        1280\,{\xi}^6\,\left( 1 + {\xi}^2 \right)  \right)  \Big\}
\nonumber\\[1.00mm]&&
     +   M^4\,\Big\{ 44\,{\chi}^{12}\,{\sqrt{1 + {\xi}^2}}
     +  176\,{\chi}^{10}\,{\xi}^2\,{\sqrt{1 + {\xi}^2}} +
     264\,{\chi}^8\,{\xi}^4\,{\sqrt{1 + {\xi}^2}}
\nonumber\\[1.00mm]&&
     +  176\,{\chi}^6\,{\xi}^6\,{\sqrt{1 + {\xi}^2}} +
     44\,{\chi}^4\,{\xi}^8\,{\sqrt{1 + {\xi}^2}} \Big\}
\Big]\,,
\eeqa
\esubeqs
where we have expanded the $R$ and $K$ numerators in powers of $M$.

\section{4D interpolating metrics}   
\label{app:4D-interpolating-metrics}

\subsection{Schwarzschild-type coordinates}   
\label{subapp:Schwarzschild-type-coordinates}

In this appendix, we present a 4D metric which interpolates
between the vacuum-defect wormhole~\cite{Klinkhamer2023-defect-WH}
at the wormhole throat ($\xi=0$) and the
Schwarzschild-defect wormhole~\cite{Wang2023}
at spatial infinity ($\xi\to\pm\infty$).
For that, we simply replace the mass parameter $M$ in
the metric \eqref{eq:4D-metricSchwarzschild-defect-wormhole}
by an appropriate function $\mu(\xi)$.

Specifically, we make the following \emph{Ansatz}:
\bsubeqs\label{eq:4D-interpolating-metric}
\begin{eqnarray}
\label{eq:4D-interpolating-metric-ds2}
ds^{2}
&=&
- \left(1-\frac{2\,\mu(\xi)}{\sqrt{b^{2}+\xi^{2}}}\right)\;dt^{2}
+
\left(1-\frac{2\,\mu(\xi)}{\sqrt{b^{2}+\xi^{2}}}\right)^{-1}\,
\frac{\xi^{2}}{b^{2}+\xi^{2}}\;d\xi^{2}
\nonumber\\[1mm]
&&
+ \left(b^{2} + \xi^{2}\right)\,
  \Big[ d\theta^{2} + \sin^{2}\theta\, d\phi^{2} \Big]\,,
\\[2mm]
\label{eq:4D-interpolating-metric-b}
b &>& 0\,,
  \\[2mm]
\label{eq:4D-interpolating-metric-mu-restriction}
2\,\mu(\xi) &<& \sqrt{b^{2}+\xi^{2}}\,,
\\[2mm]
\label{eq:4D-interpolating-metric-mu-bcs}
\mu(0) &=& 0\,,
\quad 
\lim_{\xi\to\pm\infty}\mu(\xi) = M \,,
\end{eqnarray}
\esubeqs
with $b$ the positive defect length scale  
and $\mu(\xi)$ the mass-type deformation function
(in principle, it is possible to have a regular metric with $M>b/2$).
Not surprisingly, we obtain rather bulky expressions for the
Ricci curvature scalar $R\equiv g^{\mu\nu}\,R_{\mu\nu}$
and the Kretschmann curvature
scalar $K\equiv R^{\mu\nu\rho\sigma}\,R_{\mu\nu\rho\sigma}$,
as well as for the Einstein tensor
$G^{\,\mu}_{\;\;\,\nu}\equiv
R^{\,\mu}_{\;\;\,\nu} - \half\, g^{\,\mu}_{\;\;\,\nu}\,R$.

By way of illustration, we make an explicit choice for the
interpolation function,
\beq
\label{eq:mu-tilde}
\widetilde{\mu}(\xi) = M\;\frac{\xi^{6}}{b^{6} + \xi^{6}}\,,
\eeq
where the sixth order is needed to get pure
vacuum at the wormhole throat.
The resulting Ricci curvature scalar,
\beq
\label{eq:R-tilde}
\widetilde{R}(\xi) =
\frac{12\,b^6\,M\,{\xi}^2\,\left( 4\,b^6 + 3\,b^4\,{\xi}^2 - 3\,b^2\,{\xi}^4 -
      5\,{\xi}^6 \right) }{{\left( b^2 + {\xi}^2 \right) }^{5/2}\,
    {\left( b^4 - b^2\,{\xi}^2 + {\xi}^4 \right) }^3}\,,
\eeq
is seen to vanish at the wormhole throat ($\xi=0$)
and to drop to zero faster than $|\xi|^{-3}$
towards spatial infinity ($\xi\to\pm\infty$).
The resulting Kretschmann  curvature scalar,
\beqa
\label{eq:K-tilde}
\hspace*{-10mm}
\widetilde{K}(\xi) &=&
\frac{1}
{{\left( b^2 + {\xi}^2 \right) }^5\,
{\left( b^4 - b^2\,{\xi}^2 + {\xi}^4 \right) }^6}\;
\nonumber\\[1mm]
\hspace*{-10mm}
&&
\Big[48\,M^2\,{\xi}^4\,\Big( 48\,b^{24} - 24\,b^{22}\,{\xi}^2 +
      47\,b^{20}\,{\xi}^4
- 268\,b^{18}\,{\xi}^6 + 136\,b^{16}\,{\xi}^8- 150\,b^{14}\,{\xi}^{10}
\nonumber\\[1mm]
\hspace*{-10mm}
&&
+ 347\,b^{12}\,{\xi}^{12}
-  100\,b^{10}\,{\xi}^{14} + 69\,b^8\,{\xi}^{16} - 38\,b^6\,{\xi}^{18} +
      10\,b^4\,{\xi}^{20} - 4\,b^2\,{\xi}^{22} + {\xi}^{24} \Big)\Big]\,,
\eeqa
also vanishes at the wormhole throat
and asymptotically approaches zero as $48\,M^2/\xi^{6}$.

\begin{figure}[t]   
\vspace*{-0mm}
\begin{center}
\includegraphics[width=0.60\textwidth]
{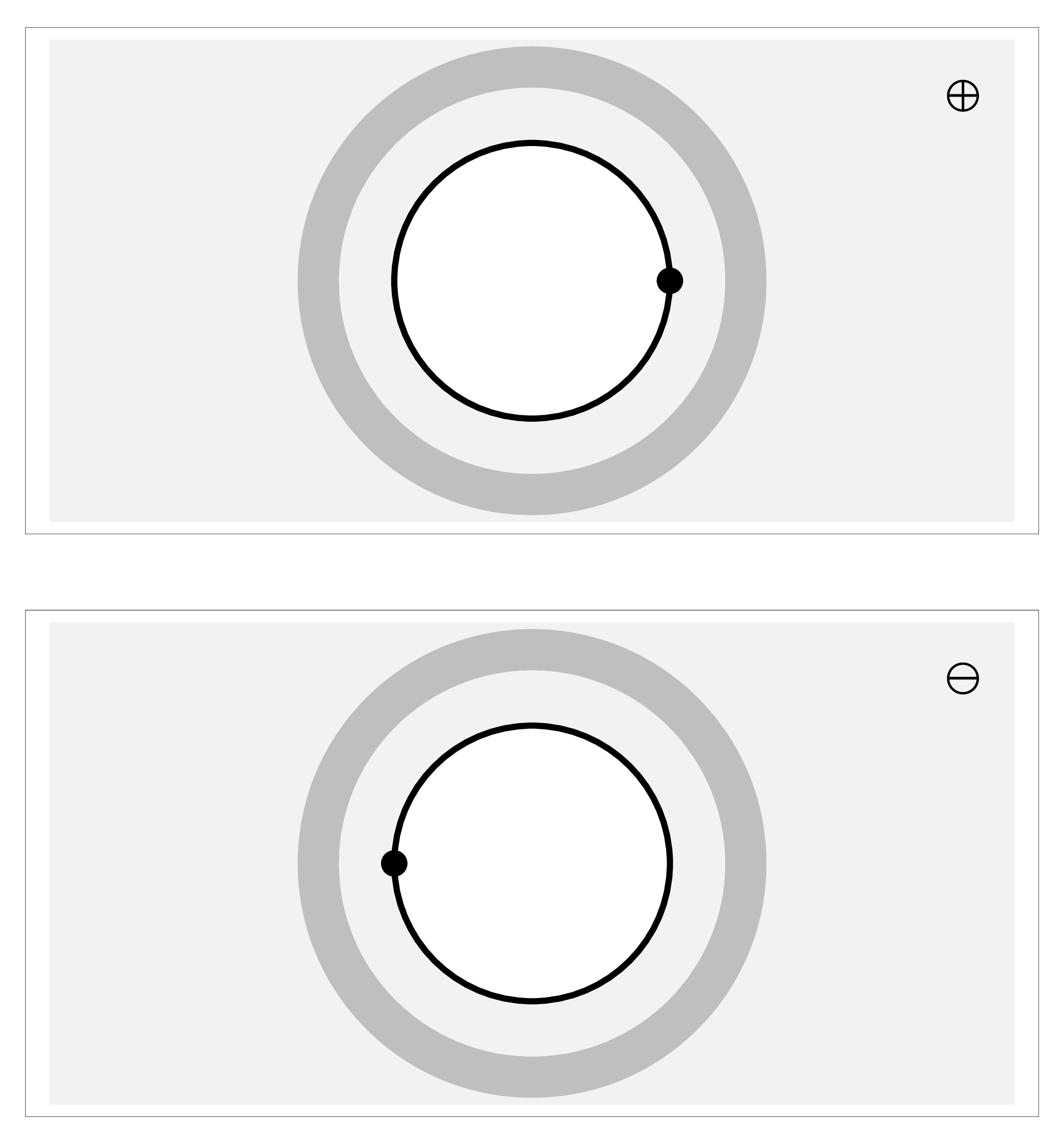}
\end{center}
\vspace*{-0mm}
\caption{Sketch of the 4D defect-wormhole 
spacetime \eqref{eq:4D-interpolating-metric}
with matter shells (shown as dark annuli),
at $\theta=\pi/2$ and equal $t=\text{const}$.
The top panel corresponds to the ``upper'' world ($\xi\geq 0$) 
and the bottom panel to the ``lower'' world ($\xi\leq 0$), 
with handedness $+/-$ as indicated in the top-right corners.
The two worlds are connected by the wormhole throat, shown as 
two heavy circles with ``antipodal'' spacetime points identified
(one spacetime point is marked by the heavy dot). 
}
\label{fig:4D-defect-wormhole-with-matter-shells}
\vspace*{-0mm}
\end{figure}

Finally, the resulting Einstein tensor is diagonal
with the following entries:
\bsubeqs\label{eq:EUd-tilde}
\beqa
\widetilde{G}^{\,t}_{\;\;\,t}(\xi) &=&
\widetilde{G}^{\,\xi}_{\;\;\,\xi}(\xi) =
\frac{-12\,b^6\,M\,{\xi}^4}
  {{\left( b^2 + {\xi}^2 \right) }^{5/2}\,
    {\left( b^4 - b^2\,{\xi}^2 + {\xi}^4 \right) }^2}\,,
\\[2mm]
\widetilde{G}^{\,\theta}_{\;\;\,\theta}(\xi) &=&
\widetilde{G}^{\,\phi}_{\;\;\,\phi}(\xi) =
\frac{-6\,b^6\,M\,{\xi}^2\,\left( 4\,b^6 + b^4\,{\xi}^2 - b^2\,{\xi}^4 -
      7\,{\xi}^6 \right) }{{\left( b^2 + {\xi}^2 \right) }^{5/2}\,
    {\left( b^4 - b^2\,{\xi}^2 + {\xi}^4 \right) }^3}\,.
\eeqa
\esubeqs
This Einstein tensor $\widetilde{G}^{\,\mu}_{\;\;\,\nu}(\xi)$ has
all entries vanishing at the wormhole throat
and asymptotically approaches zero  as $|\xi|^{-9}$,
specifically
$G^{\,\mu}_{\;\;\,\nu} \sim
b^2\,M \;|\xi|^{-9}\;
\big[\text{diag} \left(-12,\, -12,\, 42,\, 42\right)\big]^{\,\mu}_{\;\;\,\nu}$.
From the Einstein equation,
the same holds for the matter energy-momentum tensor
$T^{\,\mu}_{\;\;\,\nu}$,
up to an overall factor $1/(8\pi G)$.

The spacetime from \eqref{eq:4D-interpolating-metric}
corresponds to a defect wormhole with two matter shells
(Fig.~\ref{fig:4D-defect-wormhole-with-matter-shells})
and we have two remarks. First,
it makes sense to have a vanishing
matter energy-momentum tensor at the wormhole throat $\xi=0$, where the 
handedness flips~\cite{Klinkhamer2023-mirror-world}.
Second, it appears, in principle, possible to have different mass shells
in the upper ($\xi>0$) and lower ($\xi<0$) worlds, so that
the two asymptotic Schwarzschild-type mass parameters 
[$\lim_{\xi\to\pm\infty}\mu(\xi) = M_{\pm}$]
may differ, $M_{+} \ne M_{-}\,$.

\subsection{Painlev\'{e}--Gullstrand-type coordinates}   
\label{subapp:Painleve-Gullstrand-type-coordinates}

The 4D $M$-deformed-vacuum-defect-wormhole metric
\eqref{eq:4D-metricSchwarzschild-defect-wormhole}
from Refs.~\cite{Wang2023,Ahmed2023}
relies on Schwarzschild-type coordinates and,
therefore, has the restriction $M<b/2$. 
The mass-shell modification presented in 
App.~\ref{subapp:Schwarzschild-type-coordinates}
has a similar restriction, as given 
by \eqref{eq:4D-interpolating-metric-mu-restriction}.
We will show that these restrictions can be avoided
if we use Painlev\'{e}--Gullstrand-type 
coordinates~\cite{Painleve1921,Gullstrand1922}.

The improved 4D $M$-deformed-vacuum-defect-wormhole metric
follows from the following line element:
\bsubeqs\label{eq:4D-M-deformed-vacuum-defect-wormhole-metric}
\begin{eqnarray}
\hspace*{-5mm}
ds^{2}
&=&
- dt^{2} +
\left(\frac{\xi}{\sqrt{b^{2}+\xi^{2}}}\;d\xi 
+ \sqrt{\frac{2M}{\sqrt{b^{2}+\xi^{2}}}}\,dt\right)^{2}
+ \left(b^{2} + \xi^{2}\right)\,
  \Big[ d\theta^{2} + \sin^{2}\theta\, d\phi^{2} \Big]\,,
\\[2mm]
\hspace*{-5mm}
b &>& 0\,, \quad M \geq 0\,,
\end{eqnarray}
\esubeqs
which is a direct generalization of 
the original spacetime-defect metric
of App.~C in Ref.~\cite{Klinkhamer2014-MPLA}.
It is now possible to have $M \geq b/2$.

The corresponding mass-shell-defect-wormhole metric is
given by
\bsubeqs\label{eq:4D-mass-shell-defect-wormhole-metric}
\begin{eqnarray}
\hspace*{-5mm}
ds^{2}
&=&
- dt^{2} +
\left(\frac{\xi}{\sqrt{b^{2}+\xi^{2}}}\;d\xi 
+ \sqrt{\frac{2\,\mu(\xi)}{\sqrt{b^{2}+\xi^{2}}}}\,dt\right)^{2}
+ \left(b^{2} + \xi^{2}\right)\,
  \Big[ d\theta^{2} + \sin^{2}\theta\, d\phi^{2} \Big]\,,
\\[2mm]
\hspace*{-5mm}
b &>& 0\,, 
\\[2mm]
\label{eq:4D-mass-shell-defect-wormhole-metric-mu-restriction}
\hspace*{-5mm}
\mu(\xi) & \geq& 0\,,
\\[2mm]
\label{eq:4D-mass-shell-defect-wormhole-metric-mu-bcs}
\hspace*{-5mm}
\mu(0) &=& 0\,,
\quad 
\lim_{\xi\to\pm\infty}\mu(\xi) = M \,.
\end{eqnarray}
\esubeqs
Taking the explicit mass-deformation function \eqref{eq:mu-tilde},
the results for the Ricci curvature scalar,
the Kretschmann curvature scalar,
and the Einstein tensor are identical to
the previous expressions given by \eqref{eq:R-tilde},
\eqref{eq:K-tilde}, and \eqref{eq:EUd-tilde},
but now $M$ can be arbitrarily large compared to $b/2$.

\end{appendix}

\newpage 

\end{document}